\documentstyle[epsfig,preprint,epsf,aps]{revtex}
\tightenlines

\newcommand{\lsim}{\mathrel{\lower4pt\hbox{$\sim$}}
\hskip-12.5pt\raise1.6pt\hbox{$<$}\;}
\newcommand{\gsim}{\mathrel{\lower4pt\hbox{$\sim$}}
\hskip-12.5pt\raise1.6pt\hbox{$>$}\;}

\begin{document}

\baselineskip14pt

\vspace{-.5in}
\hskip4.5in
\vbox{\hbox{BNL--HET--98/41}
\hbox{PURD--TH--98--12}
\hbox{hep-ph/9810552}
\hbox{Revised December 1998}}

\vspace{.5in}
\begin{center}
{\Large\bf CP Violation in a two-Higgs doublet model \\ \medskip
for the top quark: \boldmath$B\to \psi K_S$ }
\vspace{.5in}

Ken Kiers$^a$\footnote{Email: kkiers@css.tayloru.edu,
$^\dag$soni@bnl.gov, $^\ddag$wu@physics.purdue.edu}, Amarjit 
Soni$^{b\dag}$ and Guo-Hong Wu$^{c\ddag}$
\vspace{.3in}

{\it $^a$Physics Department, Taylor University\\236 West Reade Ave., Upland,
IN 46989, USA}
\medskip

{\it $^b$High Energy Theory, Department of Physics\\
Brookhaven National Laboratory,Upton, NY 11973-5000, USA}
\medskip

{\it $^c$Department of Physics, Purdue University\\
        West Lafayette, IN 47907, USA}
\end{center}
\bigskip

\baselineskip24pt
\begin{abstract}
We explore charged-Higgs CP-violating effects in an intriguing two-Higgs
doublet model which accords special status to the top quark.  In this model
the heaviness of the top quark
originates naturally from the much larger VEV of the second
Higgs doublet compared to that of the first. 
The phenomenology of this model is quite distinct from that of the
usual formulations of the two-Higgs doublet model.  In particular,
the model can easily account for the observed CP violation in the
kaon sector even if the CKM matrix is real. The
associated non-standard CP phase can be
monitored through measurements of the time-dependent CP
asymmetry in $B\to \psi K_S$ in experiments at the upcoming
$B$-factories.
\end{abstract}

\newpage

Despite the stunning successes of the Standard Model (SM) it is widely
believed that it cannot be the complete theory.  It is quite possible
that the extraordinary mass scale of the top quark is giving us a hint
as to the nature of the physics beyond the SM\null. In this note we
pursue this theme within the context of an atypical two-Higgs doublet
model (2HDM) which gives the top quark a special status. The gigantic
mass scale of the top quark arises naturally in this extension 
of the SM since the
top is the only fermion receiving its mass from the vacuum expectation
value (VEV) of the second Higgs doublet. All the other fermions receive
their masses (mostly) from the VEV of the first Higgs doublet.
Besides providing a natural explanation for the largeness of
$m_t$, the introduction of
the second Higgs doublet has an important bonus in that it
allows a new  source of CP
violation: the charged Higgs sector of the
model contains a CP-violating phase (``$\delta$'')
in addition to the usual Cabibbo-Kobayashi-Maskawa (CKM)
phase \cite{kobay} of the SM.  
This phase may play a useful role in baryogenesis since the
CKM phase is believed not to be able to account for the baryon asymmetry 
of the universe in such calculations~\cite{cohen}.  Perhaps the most
exciting feature 
of the new phase $\delta$ is that it can be monitored in experiments
at the upcoming $B$ factories through time-dependent CP studies of
$B\to \psi K_S$.

The specific model which we consider was introduced in Ref.~\cite{daskao}
and is essentially a special case of the 2HDM of type III 
(i.e., Model III)~\cite{type3}. 
We regard this model as an effective low-energy 
theory in which only the (slightly extended)
Higgs sector and the regular SM particles remain as relevant
dynamical particles.  
The above-mentioned ansatz regarding the quark masses leads
to a very specific pattern of flavour- and (in general)
CP-violating couplings. 
In particular, this model exhibits tree-level flavour-changing neutral Higgs
interactions, but only in the up-type quark sector.  
It was shown in Ref.~\cite{daskao} that this model can have
significant effects on the electron electric dipole moment and
on $D$-$\overline{D}$ mixing.

The main purpose of this work is to point out that this 
top-two-Higgs-doublet model (T2HDM)
can give rise to a very significant amount 
of non-standard CP violation in charged
Higgs interactions \cite{rmk}. Contrary to conventional wisdom, perhaps, these
non-standard interactions are not particularly well-constrained by
flavour-changing neutral current processes.  For the physics of
interest to this work the important unknown parameters of the model are
$\tan\beta$ (i.e., $v_2/v_1$), $m_H$ and $\delta$. 
We have delineated the allowed parameter space of the model
by considering the
repercussions for $K$ and $B$ decays and for $K$-$\overline K$ and
$B$-$\overline B$ mixing.  Since the parameter space is relatively
large, we have chosen to focus on three ``case studies.''  In the first
we take the SM CKM phase $\gamma$ \cite{hquinn} to be identically zero.  
This scenario is particularly interesting since the CP-violating phase
$\delta$ in the Higgs  sector of the theory
is then entirely responsible for the CP violation observed in the kaon
system. In the second case study we set $\gamma$ equal  to $68^\circ$,
which is the central value of the SM fits 
(i.e., $\gamma=68^\circ\pm15^\circ$~\cite{paganini}). 
For the third case, $\gamma$ is taken to be negative and equal to
$-45^\circ$.
The important point is that, in general, the CP
asymmetry for $B\to \psi K_S$ in this model is appreciably different from the
value expected within the context of the
SM\null. One feature which distinguishes the
present analysis from many previous studies of non-standard effects in
$B\to \psi K_S$ is that in the 
present case the effect comes from the new CP-violating phase in the $B$
and  $\overline{B}$ decay amplitudes, not from a new phase in
$B$-$\overline{B}$ mixing~\cite{gngl}. 

Let us briefly recapitulate some important features of the model of
Ref.~\cite{daskao}. Consider the following Yukawa Lagrangian: 

\begin{equation}
        {\cal L}_Y  =  -\overline{L}_L\phi_1 E\ell_R
                -\overline{Q}_L\phi_1 Fd_R
                -\overline{Q}_L\widetilde{\phi}_1 G {\bf 1^{(1)}} u_R
                -\overline{Q}_L\widetilde{\phi}_2 G {\bf 1^{(2)}} u_R
                + {\rm h.c.},
        \label{eq:yuk}
\end{equation}

\noindent where $\widetilde{\phi}_i = i\sigma^2\phi_i^\ast$ $(i=1,2)$, 
and where $E$, $F$ and $G$ are $3\times3$ matrices in generation space;
${\bf 1^{(1)}}\equiv {\rm diag}(1,1,0)$;
${\bf 1^{(2)}}\equiv {\rm diag}(0,0,1)$;
and $Q_L$ and $L_L$ are the usual left-handed quark and lepton doublets.
This Lagrangian gives special status to the top
quark, as evidenced by the fact that only $\phi_2$ (and not
$\phi_1$) couples to $t_R$.  Let us set the VEVs of $\phi_1$ and $\phi_2$ to be
$v_1/\sqrt{2}$ and $v_2 e^{i\sigma}/\sqrt{2}$,
respectively.  In keeping with the spirit of
this model, we require that $\tan\beta=v_2/v_1$ be relatively
large (say at least of order twenty).

The expression for the quark-charged-Higgs Lagrangian takes the form:
 
\begin{eqnarray}
        {\cal L}^C_Y & = & (g/\sqrt{2}m_W)\left\{
                -\overline{u}_L V m_d d_R\left[G^+ -\tan\beta H^+\right]
                +\overline{u}_R m_u V d_L\left[G^+ -\tan\beta H^+\right]
                        \right. \nonumber \\
                & & \;\;\;\;\;\;\;\;\;\;\;\;\;\; \left.
                +\overline{u}_R \Sigma^\dagger V d_L\left[\tan\beta +
                        \cot\beta\right]H^+ +{\rm h.c.}\right\},
        \label{eq:chiggs}
\end{eqnarray}

\noindent where $G^\pm$ and $H^\pm$ represent the would-be Goldstone bosons
and the physical charged Higgs bosons, respectively.
Here $m_u$ and $m_d$ are the diagonal up- and down-type mass matrices,
$V$ is the usual CKM matrix and
$\Sigma \equiv m_u U_R^\dagger {\bf 1^{(2)}} U_R$.  $U^\dagger_R$ is
the unitary matrix which diagonalizes the right-handed up-type quarks. 
Since $\Sigma$ is in general not a diagonal matrix, the last
term in Eq.~(\ref{eq:chiggs}) can give rise to unusual
couplings that can violate CP
in non-standard ways.  To see this, let us write
the unitary matrix $U_R$ in a form similar to that used in
Ref.~\cite{daskao}:

\begin{equation}
U_R = \left( \begin{array}{ccc}
\cos\phi & -\sin\phi & 0 \\
\sin\phi & \cos\phi & 0 \\
0 & 0 & 1 
\end{array} \right)
\left( \begin{array}{ccc}
1 & 0 & 0 \\
0 & \sqrt{1-|\epsilon_{ct}\xi|^2} & -\epsilon_{ct}\xi^\ast \\
0 & \epsilon_{ct}\xi & \sqrt{1-|\epsilon_{ct}\xi|^2} 
\end{array} \right),
\end{equation}

\noindent where $\epsilon_{ct}\equiv m_c/m_t$ and where 
$\xi$ is a complex number
of order unity \cite{theform}.
Inserting $U_R$ into the definition of $\Sigma$, we obtain

\begin{equation}
        \Sigma = \left(\begin{array}{ccc}
         0 & 0 & 0 \\
         0 & m_c \epsilon_{ct}^2|\xi|^2 &
         m_c\epsilon_{ct}\xi^\ast\sqrt{1-|\epsilon_{ct}\xi|^2} \\
         0 & m_c\xi\sqrt{1-|\epsilon_{ct}\xi|^2} &
         m_t\left(1-|\epsilon_{ct}\xi|^2\right)
                \end{array} \right) .
\end{equation}

\noindent It is clear from this expression that this model generically
has a rather large non-standard $b$-$c$-$H$ vertex.  This can  
lead to significant CP violation effects in $B\to \psi K_S$. 
Consider, in particular, the time-dependent CP-asymmetry
$a(t)\equiv\left[\Gamma(B(t))-\Gamma(\overline{B}(t))\right]/
\left[\Gamma(B(t))+\Gamma(\overline{B}(t))\right]$, for $B\to \psi
K_S$.  In the SM $a(t)$ is free of hadronic uncertainties and is given by
$a_{\rm SM}(t) = - \sin\left( 2\beta_{\rm CKM}\right)\sin
        \left(\Delta Mt\right)$~\cite{hquinn},
where $\Delta M = M_{B_H}-M_{B_L}$ is the mass difference between
the neutral $B$ mesons and
$\beta_{\rm CKM} \equiv \arg\left(-V_{cd}V_{cb}^\ast/
V_{td}V_{tb}^\ast\right)$. Existing experimental information seems to 
constrain $\beta_{\rm CKM}$  quite tightly:
$\sin(2\beta_{\rm CKM}) = .75\pm.10$ \cite{paganini}. 
The asymmetry $a(t)$ is a
very clean way to measure the CKM angle $\beta_{\rm CKM}$ and thereby
provides an important test of  the SM \cite{subscript}.

CP-odd effects due to new physics can affect the asymmetry $a(t)$
in several qualitatively different ways.  The most direct effect comes
about if we allow the new interactions to be partly (or fully) responsible
for the observed CP violation in the kaon system.  In this case
$\beta_{\rm CKM}$ itself can be different from the value expected
within the context of the SM, and can even be
zero or negative.  For a given value of $\beta_{\rm CKM}$ there can also be new
contributions to $a(t)$ arising from $B$-$\overline{B}$ mixing and from
CP violation in the $B$ and $\overline{B}$ decay amplitudes themselves.
The  $B$-$\overline{B}$ mixing  effect, being suppressed 
 either by $1/\tan^4 \beta$ for large $\tan \beta$ or by $m_b^2/m_H^2$,  
is generically quite small in this model and may safely be ignored. 
By way of contrast, CP violation in the
amplitudes can in principle be 
appreciable in this model, leading to a sizable correction to
 the asymmetry $a(t)$ compared to the value expected within the SM.
Ignoring perturbative QCD effects, there are
two tree-level diagrams which contribute to $B\to \psi K_S$:  the SM
$W$-mediated diagram and the new charged-Higgs-mediated diagram.
The effective Lagrangian is then well-approximated by

\begin{equation}
        {\cal L}_{\rm eff} \simeq -2\sqrt{2} G_F V_{cb}V_{cs}^\ast
                \left[ \overline{c}_L\gamma_\mu b_L
                        \overline{s}_L\gamma^\mu c_L
                +2\zeta e^{i\delta}
                \overline{c}_Rb_L\overline{s}_Lc_R\right] +{\rm h.c.},
        \label{eqn:leff}
\end{equation}

\noindent where $\zeta e^{i\delta} \equiv (1/2)(V_{tb}/V_{cb})
(m_c\tan\beta/m_H)^2\xi^\ast$  with $\zeta$ taken to be real and positive.
In writing the above expression, we have
neglected terms which are subdominant for $\tan\beta\gg 1$ and
$|\xi|={\cal O}(1)$.  Note that the Higgs-mediated
contribution can in principle be large since it is proportional
to $V_{tb}/V_{cb}$$\sim$$25$ as well as to $\tan^2\beta$ \cite{rate}.

Using Fierz identities and assuming factorization, we may write
the total amplitude for $B\to\psi K_S$ as
${\cal A} \equiv {\cal A}(B\to\psi K_S) \simeq
                {\cal A}_{\rm SM}
                        \left[ 1 - \zeta e^{-i\delta}\right].$
To the extent that factorization holds, there is no relative
strong phase between the $W$- and charged-Higgs-mediated
diagrams, so that the $B$ and $\overline{B}$ decay
amplitudes have the same
magnitude and differ only by a CP-violating phase.
We may thus write
$\overline{\cal A}/{\cal A} = \left(\overline{\cal A}_{\rm SM}/
        {\cal A}_{\rm SM}\right)\exp(- 2 i \vartheta),$
where $\vartheta$ represents the correction to $\beta_{\rm CKM}$ due
to the Higgs-mediated contribution to $B\to \psi K_S$; i.e.,
the time-dependent asymmetry $a(t)$
now measures $\beta_{\rm CKM}+\vartheta$
instead of $\beta_{\rm CKM}$.  The angle $\vartheta$ is simply given
 by $\tan\vartheta = \zeta\sin\delta/ (1-\zeta\cos\delta)$ and can,
in principle, be rather large. 

Let us discuss some of the experimental constraints on $\vartheta$.
The most stringent experimental constraints on the parameter
space of the T2HDM come from $b\to s \gamma$ and
$K$-$\overline{K}$ mixing. $b\to s\gamma$ places an important
constraint on the parameter space of the 
T2HDM just as it does for  Model II, where a $\tan\beta$-independent
lower bound of approximately $370$ GeV may be placed on the charged Higgs
mass~\cite{ciuchini}.  The situation in the present case is
somewhat complicated by the fact that the Higgs contribution
in this model has a CP-violating phase.
This extra degree of 
freedom means that the new contribution to $b\to s\gamma$ is
complex and need not interfere constructively
with the SM contribution at LO, as is the case in Model II\null.
As a result, the shape of the excluded region in the $\tan\beta$-m$_H$
plane is strongly dependent on the CP-violating phase $\delta$.

In order to minimize the theoretical uncertainties associated
with $b\to s\gamma$,
it is customary to consider the ratio $R = {\cal B}(b\to X_s \gamma)/
                {\cal B}(b\to X_c e\overline{\nu})$.
The branching ratio appearing in the numerator of this expression
has been investigated by both the CLEO II
and ALEPH collaborations \cite{cleo2}.  The experimental results
may be combined to obtain the weighted average
$R^{\rm exp} = \left(2.90\pm0.46\right)\times 10^{-3}.$
Considerable progress has also been made on the theoretical side
of this decay in recent years.  In particular, complete NLO QCD
corrections~\cite{chetyrkin}
as well as some two-loop electroweak corrections~\cite{czarmarc}
have now been incorporated into the calculation of $R$ within
the context of the SM~\cite{loeffect}.

The T2HDM can affect
the ratio $R$ through diagrams similar to the usual SM
diagrams~\cite{inamilim} in which the $W$ boson is replaced
by a charged Higgs. 
The Higgs contribution gives a correction to the matching
condition for the LO Wilson coefficient $C_7^{(0)}$ at the scale $m_W$
so that the relevant effective Wilson coefficient
at the scale $\mu \sim {\cal O}(m_b)$
is modified according to
$C_7^{(0)\rm eff} (\mu)\rightarrow C_7^{(0)\rm eff}(\mu)+
 (\alpha_s(m_W)/\alpha_s(\mu))^{16/23} \delta C_7^{(0)}(m_W),$ where

\begin{eqnarray}
        \delta C^{(0)}_7(m_W)\!\! & = &\!\!\! \sum_{u=c,t}\kappa^u
                \left[ -\tan^2\beta +
                \left(\Sigma^T V^\ast\right)_{us}\left(\tan^2\beta +1
  \right)/ m_u V_{us}^\ast \right]  \label{eq:delc7} \\
               \!\! & &\!\!\! \times\left\{B(y_u) + A(y_u)
                        \left[-1 +
                \left(\Sigma^\dagger V\right)_{ub} \left(\cot^2\beta + 
 1\right)/ m_u V_{ub} \right]/6 \right\} . \nonumber
\end{eqnarray}

\noindent In this expression $\kappa^u = \pm 1$ for $u=c,t$;
 $y_u = (m_u/m_H)^2$; and
$A$ and $B$ are standard expressions~\cite{somebody,scalar}.

In order to compare our theoretical calculation of $R^{\rm theory}$ with
the experimental number quoted above, we assign a theoretical
uncertainty of $\delta R^{\rm theory} = 0.7\times 10^{-3}$ 
to $R^{\rm theory}$ \cite{ciuchini}.
This uncertainty corresponds roughly to the variation of $R^{\rm theory}$
 as $\mu$ is varied between $m_b/2$ and $2m_b$.
Adding the experimental and theoretical uncertainties in quadrature,
we find the following constraint on $R^{\rm theory}$: $ 0.0012 \leq
R^{\rm theory} \leq 0.0046$. 

We next consider the implications of the T2HDM
for $K$-$\overline K$ mixing. 
The total short distance contribution to $\Delta m_K$ is given by

\begin{equation}
(\Delta m_K)_{SD}  = (G_F^2/6\pi^2) f_K^2 B_K m_K \lambda_c^2
\times \left[ m_c^2 \eta_1 + (m_c^4 \tan^4 \beta/4 m_H^2) \eta^{\prime}_1 
		\right]
\end{equation}

\noindent where the first term is the usual SM contribution and the second
term is the dominant contribution in the T2HDM due to the $HHcc$ box
diagram.  Here $B_K$ is the usual bag factor,
$\lambda_c=V_{cs}V_{cd}^*$, and $\eta_1$ and 
$\eta^{\prime}_1$ are the QCD corrections to the two box diagrams.
The SM top quark contribution is a few percent of the charm quark
contribution and is not included here. Similarly,
the contributions from $WHcc$ and other box diagrams are negligible
in the large $\tan \beta$ limit and are ignored.

In order to numerically deduce the allowed parameter space subject to
the $\Delta m_K$ constraint we use the method
described in \cite{paganini}.  Assuming the magnitude of the long distance
contribution to $\Delta m_K$ to be no larger than $30\%$, we find
the $95\%$ C.L. limit $m_H/\tan^2 \beta  >  0.48$ GeV
 for $\tan \beta > 10$~\cite{errordist}.  Note that the
$\tan^4 \beta$ enhancement in $\Delta m_K$ generally leads to a very severe
lower bound on the Higgs mass for large $\tan \beta$ -- this is
 a unique feature of the T2HDM\null.

  Due to the $\tan^4 \beta$ dependence,
  the $CP$-violating parameter $\epsilon_K$ also receives its largest
correction from the $HHcc$ box diagram in the large $\tan \beta$ limit.
The SM contribution is well known \cite{buras2}.  The
dominant Higgs contribution is given by 
\begin{equation} \label{eq:epsK}
\epsilon_K^H  =  e^{i
\frac{\pi}{4}} C_{\epsilon} B_K A \lambda^4 
 \eta_1^{\prime} \sqrt{\rho^2 + \eta^2} \sin(\gamma + \delta) |\xi|
  (m_c \tan \beta)^4/4m_W^2 m^2_H
\end{equation}
\noindent where $A$, $\lambda$, $\rho$, and $\eta$ are the CKM parameters 
in the Wolfenstein parameterization \cite{wolfen},
$\gamma \equiv \tan^{-1} \eta/\rho$ is the CKM phase \cite{hquinn}, and
$C_{\epsilon}=G^2_F f^2_K m^2_W m_K/6 \sqrt{2} \pi^2 \Delta m_K
=3.78 \times 10^4$.

  Since $\gamma$ is essentially a free parameter in this model~\cite{gamnote},
we can obtain bounds on the parameter
$Y \equiv  \sin (\gamma+\delta) |\xi|
( \tan \beta/20)^4( 200 \; \mbox{GeV}/m_H)^2$
for any given value of $\gamma$
by allowing $\sqrt{\rho^2 + \eta^2}$ to vary within its $1\sigma$
uncertainties derived from $b\rightarrow u e \nu$.  (We also allow
$A$, $\eta_1^\prime$, etc. to vary, as above.)  For
$\gamma=0^\circ$ we obtain the $95\%$ C.L bound
$0.08 <  Y < 0.39$; the bound becomes 
$0.14 <  Y < 0.65$ for $\gamma=-45^\circ$. 
  If we assume that $\gamma$ takes its SM central value of $68^\circ$
\cite{paganini}, the bound becomes $-0.085 < Y  < 0.08$.
Unlike the constraint on $(m_H, \tan \beta)$ coming
from $\Delta m_K$, the one coming from $\epsilon_K$
depends on $|\xi|$ and $\delta$.

Figure~1  shows plots in the $\tan\beta$-$m_H$ plane
for a few representative values of the parameters of this model.  
In all of these plots we have fixed $|\xi|$ to be unity, which is
a natural choice in this model.
The shaded bands in Fig.~1 correspond to the regions
{\it allowed\/} by all three constraints.
The amplitude of the time-dependent
CP asymmetry in $B\to\psi K_S$ is now given
by $a_{\psi K_S} \equiv \sin\left[2(\beta_{\rm CKM}+\vartheta)\right]$.
Contours of constant $\vartheta$ have a very simple form
in the $\tan\beta$-$m_H$ plane:  they are lines of constant slope
emanating from the origin of that plane.  We have suppressed such contours 
to avoid overcrowding the plots.
Figure~2  shows the allowed values of 
$\sin\left[2(\beta_{\rm CKM}+\vartheta)\right]$ as a function of
the CP-violating parameter $\delta$.  
In order to clearly illustrate charged Higgs effects,
we fix $\sqrt{\rho^2+\eta^2}=0.41$ (this is its central value
\cite{paganini}) so that $\beta_{\rm CKM}$ is uniquely determined
for a given $\gamma$.
 This situation might well correspond
to a future scenario in which measurements of $b\to u e \nu$ have
become more precise.  

For illustration let us focus on three representative
choices for $\gamma$ (see Fig.~2).  
The first corresponds to a real CKM matrix; i.e.,
$\gamma=0^\circ$.  In this case
the phase $\delta$ is solely responsible for the observed CP violation
in the $K$-$\overline{K}$ system.
For a given $\delta$, the upper limit of the allowed region 
 corresponds to the intersection 
of the $b\to s\gamma$ curve with the more
stringent of the $\Delta m_K$ and the $\epsilon_K$ curves
in the $\tan\beta$-$m_H$ plane.
The excluded regions for $\delta$ come dominantly from the $\epsilon_K$
constraint.
It is interesting to note that small values ($\sim 5^{\circ}$) for $\delta$
are not ruled out by $\epsilon_K$ due to the $\tan^4\beta$ enhancement factor
in Eq.~(\ref{eq:epsK}). 
This scenario gives a very clear signature:
instead of measuring an asymmetry of order
$\sin 2\beta_{\rm CKM}\sim 0.75$ as expected in the SM, 
one would find an asymmetry no larger than about $0.27$.

Figure~2 also shows the allowed range of the $CP$ asymmetry
$a_{\psi K_S}$
for another phenomenologically interesting case, $\gamma=68^\circ$.
It is evident from this plot that 
significant deviations from the pure SM value 
(i.e.\ the horizontal line  corresponding
to $\vartheta=0$) are possible in this scenario. 
Depending on the value of $\delta$, the partial rate 
asymmetry in $B\to \psi K_S$ can
range between $\sim 42$--90\% and therefore may be appreciably
different from the SM expectation of $\sim 75\%$.

  As the third scenario, we show in Fig.~2 how the $CP$ asymmetry 
$a_{\psi K_S}$ can even have an opposite sign relative to the standard model
expectation. This generally occurs when $\gamma$ takes negative
values which in turn gives a negative $\beta_{\rm CKM}$.
As an example, the allowed region in the $a_{\psi K_S}$-$\delta$
plane is shown for $\gamma=-45^\circ$. 
This scenario gives a distinct signal
for physics beyond the SM.

To summarize, we have presented a study of CP violation in a 
two-Higgs-doublet model that accords a special status to the top
quark and accommodates the remarkable heaviness of the top quark rather
naturally.  
While the model has repercussions for numerous experiments \cite{edm}, 
we have focused here on its effects on the time-dependent
CP asymmetry in $B\to \psi K_S$.  This asymmetry is of particular
interest due to the fact that reliable theoretical 
predictions can be made here and also because intense experimental activity 
by the $B$ factories is anticipated on this decay mode in the very near future.
The asymmetry in this model can differ significantly from the predictions
of the SM and can even change sign with respect to the SM expectation.
These features may be helpful in leading to the discovery of new physics.

We thank D. Atwood, A. Czarnecki and O. Vives for helpful
discussions and comments.
This research was supported in part by the U.S. Department of Energy
under contract numbers  DE-AC02-98CH10886 and DE-FG02-91ER40681 (Task B).
K.K. was also supported in part by the Natural Sciences and
Engineering Research Council of Canada and by an SRTP
grant at Taylor University.  G.W. is grateful for the hospitality
of the High Energy Theory Group at Brookhaven National Lab and of
the Fermilab Theoretical Physics Department.

\begin{figure}[hbt]
\centerline{\epsfig{figure=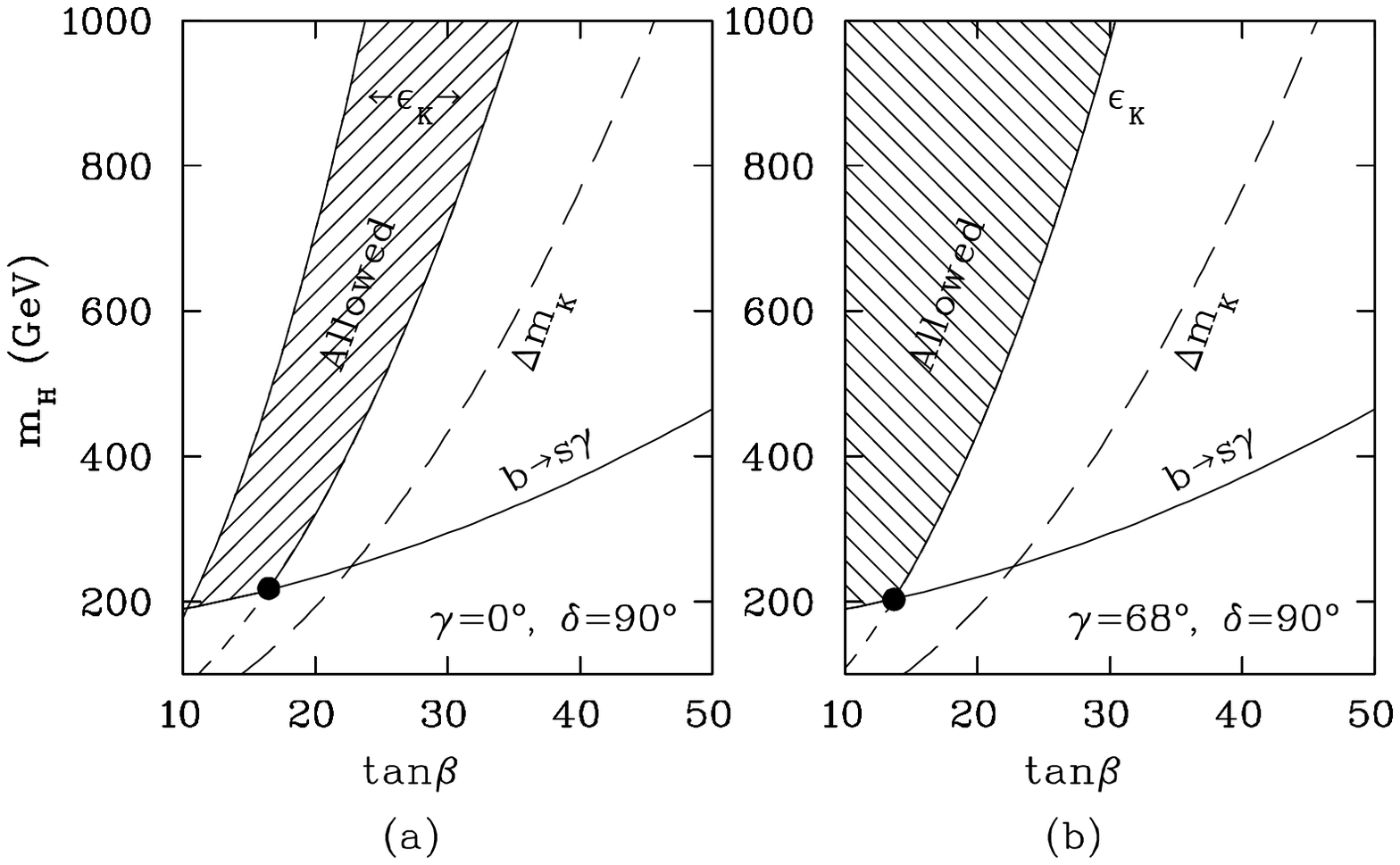}}
\medskip
\medskip
\caption{Experimental constraints on the T2HDM from $b\to s \gamma$, 
$\epsilon_K$ and $\Delta m_K$.  The allowed regions are shaded.  The
dots indicate where the largest values of $|\vartheta|$
occur: these are 
$\vartheta=7.0^\circ$ and $\vartheta=5.5^\circ$ in (a) and (b), respectively.
The plot for $\gamma=-45^\circ$  is similar to that of $\gamma=0^\circ$,
and is not shown.}
\label{fig1}
\end{figure}

\begin{figure}[hbt]
\centerline{\epsfig{figure=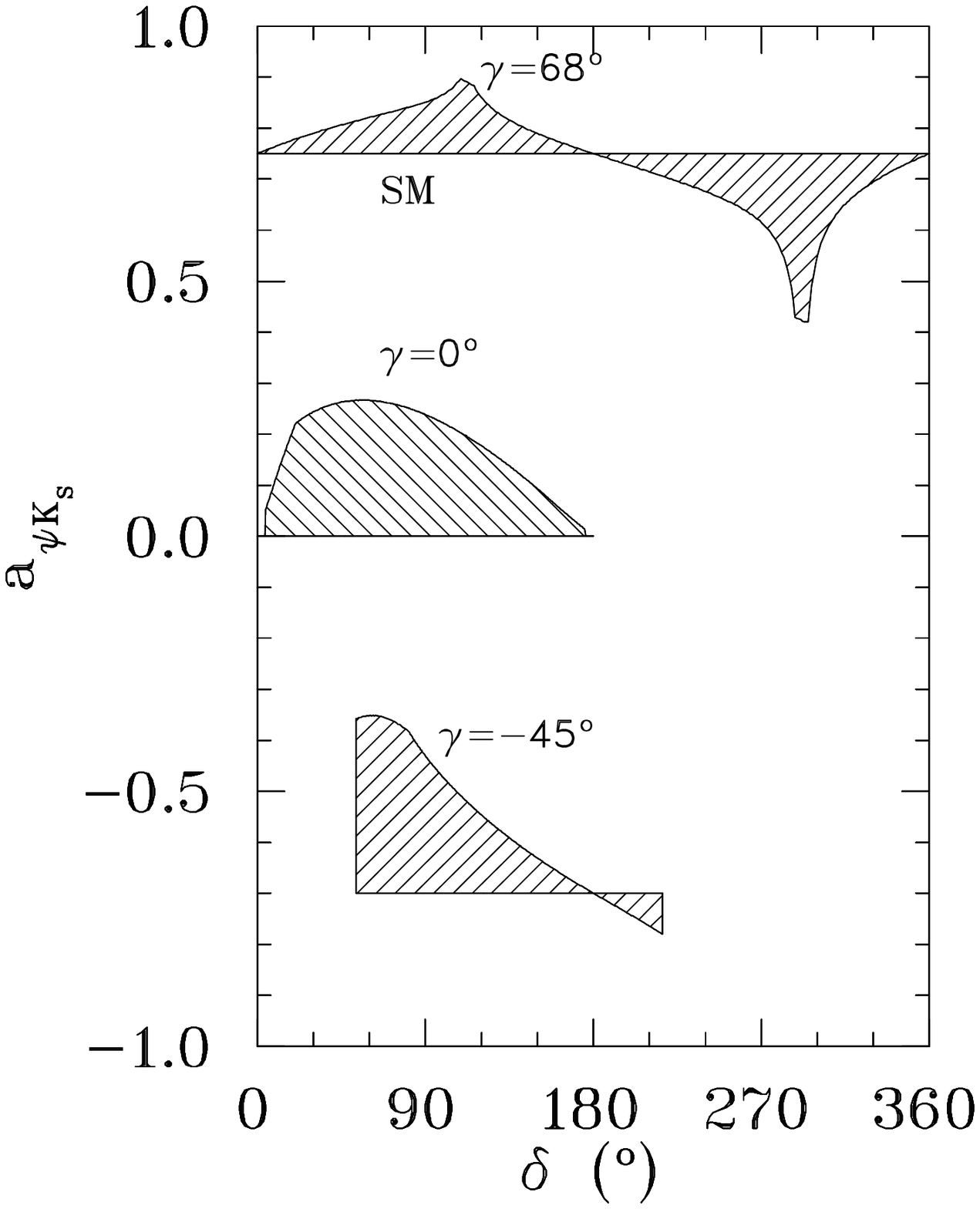}}
\medskip
\medskip
\caption{Amplitude of the time-dependent CP asymmetry 
$a_{\psi K_S} \equiv \sin (2\beta_{\rm CKM} + 2 \theta)$
  versus the non-standard 
CP-odd phase $\delta$ for $B\to\psi K_S$.
The top horizontal line is for the SM    
assuming the best fit value ($\sin 2\beta_{\rm CKM}=0.75$) of
Ref.~[6].  The shaded regions correspond to the allowed ranges of the
asymmetry in the T2HDM for three representative       
choices of $\gamma$: $\gamma=68^\circ$ is the best fit of
Ref.~[6], $\gamma=0^\circ$ corresponds to a real CKM matrix, and 
$\gamma=-45^\circ$ changes the sign of $a_{\psi K_S}$ relative to 
the SM expectation.}
\label{fig2}
\end{figure}

\begin{thebibliography}{99}
\bibitem{theform} The form which we have chosen for
$U_R$ differs slightly from that used in Ref.~\cite{daskao}, the main
difference being that we neglect possible mixing between the first and third
generations.  If this mixing is suppressed by $m_c/m_t$ or more, then
it has a negligible effect in the processes which we will consider.  Dropping
this term also allows us to write $U_R$ in a manifestly unitary form.
\bibitem{subscript} The subscript is added to $\beta_{\rm CKM}$ to
distinguish it from $\beta\equiv\tan^{-1} (v_2/v_1)$.  
\bibitem{rate} 
Indeed this non-standard source may contribute $\sim 20\%$
to the observed branching ratio for $B \rightarrow \psi K_S$
which is much smaller than the intrinsic theoretical uncertainties
in the absolute rate calculation.
\bibitem{loeffect} 
Since we restrict ourselves to computing the LO
effect of the new Higgs interactions, for consistency, we also
incorporate only the LO SM contribution to $b \to s \gamma$.  
\bibitem{scalar} Note
that we are neglecting here the scalar operator
${\cal O}_S = \overline{c}_Rb_L\overline{s}_Lc_R$, which receives
contributions at tree-level in this model [see Eq.~(\ref{eqn:leff})].
This term could potentially contribute to
$C_7^{(0)\rm eff}(\mu)$ through operator mixing
and our neglect of it is an approximation.
Analogous terms are also present in Models I and II, but are safely
neglected there because the Wilson coefficient $C_S(m_W)$ is generally
very small in these models.   
\bibitem{errordist} We take all error distributions
(both for the theoretical and the experimental quantities)
to be rectangular, with widths $\pm 1\sigma$.   
\bibitem{gamnote}
Note that the choice $\gamma=0$ is disfavored at the $1\sigma$ level
by the combined constraints from $B$-$\overline{B}$ mixing and 
$b \rightarrow u e \nu$~\cite{paganini}.  
However, to consider the full range of possibilities for this model
-- and given the theoretical uncertainties implicit
in these constraints -- we choose to adopt a rather conservative 
approach with respect to their interpretations.
\bibitem{edm}
For example, the charged Higgs exchange contribution to the neutron 
EDM in this model can be as large as $10^{-27} - 10^{-26}\; e\;\rm{cm}$.
\end{thebibliography}
\end{document}